\documentclass[12pt,twoside]{article}
\usepackage{fleqn,espcrc1,epsfig,wrapfig}
\def\apm#1{\hbox{$\pm #1$}}
\def\epm#1#2{\hbox{${\lower1pt\hbox{$\scriptstyle +#1$}}
  \atop {\raise1pt\hbox{$\scriptstyle -#2$}}$}}
\def\GeV{{\rm GeV}}
\def\PR{{\it Phys.~Rev.~}}

\def\NP{{\it Nucl.~Phys.~}}

\def\PL{{\it Phys.~Lett.~}}

\def\ZP{{\it Zeit.~Phys.~}}

\def\APP{{\it Acta.~Phys.~Pol.~}}
\def\vol#1{{\bf #1}}\def\vyp#1#2#3{\vol{#1} (#2) #3}
\def\as{\alpha_S}
\def\be{\begin{equation}}
\def\ee{\end{equation}}
\def\bea{\begin{eqnarray}}
\def\eea{\end{eqnarray}}
\def\beq{\begin{equation}}
\def\eeq{\end{equation}}
\title{Polarized parton distributions in perturbative QCD}

\author{G. Ridolfi \address{INFN Sezione di Genova \\ 
        Via Dodecaneso 33, I-16146 Genova, Italy}}
\begin{document}

\maketitle

\begin{abstract}
We review the main results of next-to-leading order QCD analyses
of polarized deep-inelastic scattering data,
with special attention to the assessment of theoretical uncertainties.
\end{abstract}

\bigskip
{\it Talk given at the Nucleon '99 Workshop, Frascati (Italy), June 7-9 1999} 
\bigskip

\section{} Experimental information on deeply inelastic scattering (DIS) of
polarized leptons off polarized nucleon targets has greatly improved
since the first E80/E130 experiments at SLAC (see ref.~\cite{ABFR2}
for a complete bibliography.) While at an early stage the attention
was mainly focused on the test of the Ellis-Jaffe~\cite{EJ} sum rule,
more recently the interest is concentrated on the study of the general
features of polarized nucleons in the deep-inelastic region in the
context of the QCD-improved parton model. The analysis of polarized
DIS data can at present be performed using perturbative QCD at
next-to-leading order accuracy, thanks to the recent
computation~\cite{NLO} of order $\as^2$ Altarelli-Parisi splitting
functions in the polarized case. This analysis has been performed by
many different authors~\cite{ABFR2},\cite{stratmann}-\cite{AB} with
consistent results. I will present here the results obtained in
ref.~\cite{ABFR2}.
The strategy is the same as the one adopted in the case of unpolarized
DIS: the polarized parton distributions $\Delta q(x,Q^2), \Delta
g(x,Q^2)$ at an initial scale $Q_0$ are assumed to have an arbitrarily
chosen $x$ dependence, specified by a set of unknown parameters; with
the help of the QCD-improved parton model formulas and of
Altarelli-Parisi evolution, one computes the structure function
$g_1(x,Q^2)$ at each data point in the $(x,Q^2)$ plane, and fits the
unknown parameters, which are in turn related to interesting physical
quantities.

A first important point is the test of the Bjorken sum rule~\cite{Bj}.
The combination
\begin{equation}
\Gamma_{Bj}\equiv\int_0^1 dx \left[g_1^p(x,Q^2)-g_1^n(x,Q^2)\right]
\end{equation}
can be shown to be proportional to the axial charge 
\beq 
g_A=\int_0^1 dx \left[\Delta u(x,Q^2) - \Delta d(x,Q^2)\right],
\eeq 
which is $Q^2$-independent because of current conservation, times a
Wilson coefficient $C_{NS}(\alpha_S)$, which is perturbatively
computable, and known to order $\alpha_S^3$. This is an important and
accurate theoretical prediction, since corrections to it may only
come from isospin violation ($\sim 1$\%), from terms of order $\as^4$
or higher in the Wilson coefficient, or from non-perturbative
contributions, suppressed by powers of $\Lambda_{QCD}^2/Q^2$.  
A direct test of the Bjorken sum rule has become possible, since 
$g_1$ data with deuteron and neutron targets are available.
This is done by using the non-singlet axial charge $g_A$ as one of the
parameters of the fitting procedure. The result~\cite{ABFR2} is
\beq
g_A=1.18\pm0.05{(\rm exp)}\pm 0.07{(\rm th)}=1.18\pm0.09,
\label{gafit}
\eeq
to be compared with the value $g_A=1.257\pm0.003$ measured in $\beta$
decay (we will discuss in the next section the unceretainties in
eq.~(\ref{gafit}). It can be concluded that polarized DIS data are
consistent with the Bjorken sum rule at the level of one standard
deviation, with an accuracy of less than 10~\%.

A second interesting question is the singlet contribution to the first
moment of $g_1$:
\beq
\left[\int_0^1 dx g_1(x,Q^2)\right]_{singlet}
= C^{(1)}_S(\alpha_S) a_0(Q^2),
\eeq
where $C_S^{(1)}(\alpha_S)$ is the first moment of the singlet
coefficient function, and $a_0(Q^2)$ the singlet axial charge; $a_0$
is {\it not} scale independent because of the axial current anomaly.
In the QCD-improved parton model, one can choose the factorization
scheme~\cite{AB} so that the first moment of the singlet combination of
polarized quark densities, $\Delta\Sigma(1)$, is scale independent,
and can therefore be interpreted as the total helicity carried by
quarks. In this class of schemes one has
\beq 
a_0(Q^2)=\Delta \Sigma(1)-n_f\frac{\alpha_s(Q^2)}{2\pi} \Delta g(1,Q^2),
\label{singlet} 
\eeq
where $\Delta g(1,Q^2)$ is the first moment of the polarized gluon
density. The values of $a_0$, $\Delta\Sigma(1)$ and $\Delta g(1,Q^2)$
can be extracted from the fitting procedure outlined above. We find
\bea
\Delta \Sigma (1) &=& 0.46\pm0.04~{\rm (exp)}\pm0.08~{\rm (th)}
= 0.46\pm0.09,
\nonumber\\
\Delta g(1,1\,{\rm GeV}^2)&=&
1.6\pm0.4~{\rm (exp)}\pm0.8~{\rm (th)} = 1.6\pm0.9,\label{firstmom}\\
a_0(\infty)&=&0.10\pm0.05~{\rm (exp)}\epm{0.17}{0.10}~{\rm (th)}
= 0.10\epm{0.17}{0.11}.
\nonumber
\eea
The results in eqs.~(\ref{firstmom}) show that large values
of $\Delta\Sigma(1)$ are compatible with small values of
$a_0$, provided $\Delta g(1,Q^2)$ is positive and large
enough, as first suggested in refs.~\cite{altarelliross}.

Finally, one can attempt using the value of $\alpha_S$ as one of the
parameters in the fit~\cite{ABFR}, as customary in unpolarized data
analyses. It is interesting to note that the value obtained with
polarized DIS data, namely
\beq
\alpha_S(m_Z)=0.120\epm{0.004}{0.005}{\rm (exp)}
\epm {0.009}{0.006}{\rm (th)}=0.120\epm{0.010}{0.008},
\label{alphafit}
\eeq 
is very close to other determinations, and that the uncertainty is
reasonably small.

\section{} 
\begin{center}
\begin{table}
\vspace*{0.5cm}
\begin{center}
\begin{tabular} {|l|l|l|l|l|l|l|}
\hline  & $g_A$ &$\Delta\Sigma$& $\Delta g$
&$ a_0$ & $\alpha_s$\\
\hline
experimental  & \apm0.05 & \apm0.04 & \apm0.4 & \apm0.05 &
\epm{0.004}{0.005} \\
\hline
fitting       & \apm0.05 & \apm0.05 & \apm0.5 & \apm0.07 & \apm0.001 \\
$\alpha_s$ \& $a_8$ & \apm0.03 & \apm0.01 & \apm0.2 &
\apm0.02 & \apm0.000 \\
thresholds    & \apm0.02 & \apm0.05 & \apm0.1 & \apm0.01 & \apm0.003 \\
higher orders   & \apm0.03 & \apm0.04 & \apm0.6 &
\epm{0.15}{0.07} & \epm{0.007}{0.004} \\
higher twists       & \apm0.03 &  -   &  -  &  -  & \apm0.004 \\
\hline
theoretical         & \apm0.07 & \apm0.08 & \apm0.8 &
\epm{0.17}{0.010} &\epm{0.009}{0.006} \\
\hline
\end{tabular}
\end{center}
\caption{Contributions to the errors in the determination of the quantities
$g_A$, $\Delta\Sigma(1)$, $\Delta g(1,1\GeV^2)$, $a_0(\infty)$ and
$\alpha_s(m_Z)$ from the fits described in the text.}
\label{errors}
\end{table}
\end{center}
\begin{wrapfigure}{r}{8truecm}
\epsfxsize=8truecm\hfil\epsfbox{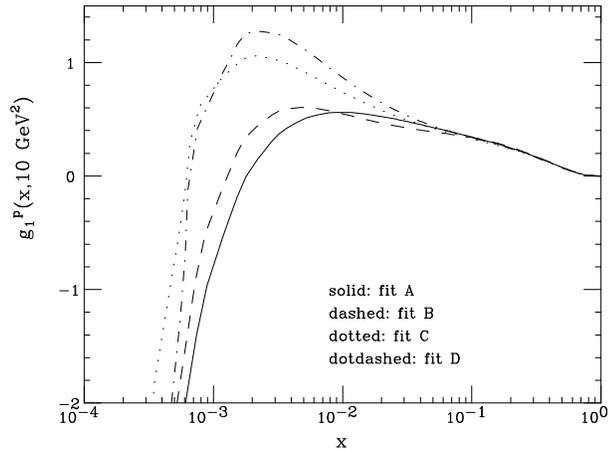}\hfil
\caption{Plot of $g_1^P(x,Q^2)$ at $Q^2=10$~GeV$^2$
for different parton density parametrizations (labelled A--D).}
\label{smallx}
\end{wrapfigure}
We come now to a discussion of the theoretical uncertainties attached
to the observables mentioned above, and summarized in
table~\ref{errors}. This analysis has been presented in
ref.~\cite{ABFR}, where the interested reader can find more details.
The experimental error is taken into account by the minimum-square
fitting procedure. We have added in quadrature systematic and
statistic errors on each data point; this procedure results probably
in an overestimate of the effective uncertainty on the fit parameters,
since it does not account for correlations among systematics.

A source of theoretical uncertainty which is often neglected is the
arbitrariness in the choice of the functional form in $x$ for the
parton densities at the initial scale. We have considered a wide range
of functional forms (see ref.~\cite{ABFR} for details), and we have
found the choice of the initial-scale parametrization affects
consideralby the final results. In fact, different parametrizations
lead to different estimates of contribution to the first moment of
$g_1$ from the small-$x$ region, where the experimental information is
very poor. The corresponding spread in the determination of physical
quantities must be included in the total uncertainty. This is
illustrated in fig.~\ref{smallx}, where the different curves refer to
different parametrizations of the initial scale parton densities. The
curves are quite close to each other in the measured region, as
expected, since they all correspond to fits of comparable quality,
while they differ considerably below $x\sim 0.01$. Perhaps, part of this
uncertainty could be reduced using positivity constraints~\cite{AFR}.

Our analysis includes data points at $Q^2$ down to 1~GeV$^2$, in order
to have a reasonable information at small values of $x$.  At such low
scales, one should worry about the uncertainty originated by higher
orders in the perturbative expansion. These can be estimated by
varying the values of renormalization and factorization scales
$\mu_R,\mu_F$ independently around $Q^2$. Not surprisingly, this turns
out to be the most important origin of theoretical uncertainty.  It
should be poited out that this ``theoretical'' uncertainty could
eventually be removed, if more data at small $x$ and higher $Q^2$ were
available; in fact, one could then exclude from the analysis data
points below, say, $Q^2\sim 4-5$~GeV$^2$, as in most unpolarized DIS
analyses, thus avoiding the region where the QCD perturbative
expansion is less reliable.

Non-perturbative contributions are also potentially large at this low
values of $Q^2$, since they have the form of powers of
$\Lambda_{QCD}^2/Q^2$. Unfortunately, they are very difficult to
estimate. One possible strategy is that of comparing results
obtained fitting all data above $Q^2=1$~GeV$^2$ with those obtained by
excluding data points below $2$~GeV$^2$. This procedure indicates that
the contribution of power-suppressed terms is not very large,
compared to other sources of uncertainty. A similar conclusion
is obtained by fitting the Bjorken sum to its perturbative expression,
supplemented with a twist-4 term $a/Q^2$, with the parameter $a$ taken
from renormalon and sum rule estimates. 

Other minor sources uncertainties, such as violations of the $SU(3)$
flavour symmetry or the position of heavy quark thresholds in $Q^2$
evolution, are also listed in table~\ref{errors}.

\bigskip 

I wish to thank G. Altarelli, R. Ball and S. Forte for the
fruitful collaboration on the subject of this talk.

\end{document}